\documentclass[%
 reprint,
 amsmath,amssymb,
 aps,
]{revtex4-2}

\usepackage{graphicx}
\usepackage{dcolumn}
\usepackage{bm}
\usepackage{booktabs}
\usepackage[table,xcdraw]{xcolor}
\usepackage{colortbl}

\usepackage[utf8]{inputenc}
\usepackage[T1]{fontenc}
\usepackage{mathptmx}
\usepackage{etoolbox}

\usepackage{graphicx}    
\usepackage{dcolumn}     
\usepackage{bm}          
\usepackage{amsmath}     
\usepackage{amssymb}     
\usepackage{multirow}    
\usepackage{color}       
\usepackage{soul}
\usepackage{enumerate}

\makeatletter
\def\@email#1#2{%
 \endgroup
 \patchcmd{\titleblock@produce}
  {\frontmatter@RRAPformat}
  {\frontmatter@RRAPformat{\produce@RRAP{*#1\href{mailto:#2}{#2}}}\frontmatter@RRAPformat}
  {}{}
}%

\newcommand{\Eq}  [1]    {Eq.~(\ref{#1})}       

\newcommand{\Fig} [1]    {\textcolor{green}{Fig.~\ref{#1}}}
\newcommand{\Figs}[1]    {Figs~\ref{#1}}        

\newcommand{\ie}         {\textit{i.e.,} }       
\newcommand{\rhs}      {r.h.s.}               

\newcommand{\dt}					{\Delta t}
\newcommand{\kBT}					{k_B T}
\newcommand{\kB}					{k_B }

\newcommand{\energy}				{\varepsilon}

\newcommand{\ALaMu}			   {A_{\mu}}

\newcommand{\AAttr}				{A_\textrm{attr}}
\newcommand{\aAttr}				{a_\textrm{attr}}
\newcommand{\bAttr}				{b_\textrm{attr}}

\newcommand{\Sphere}       {\text{S}}
\newcommand{\Rod}          {\text{R}}

\newcommand{\asyn}              {$\alpha$-syn}

\newcommand{\cMod}			   {\mathrm{c_M}}
\newcommand{\constAttr}				{\mathcal{C}}
\newcommand{\constRed	}			{\constAttr_{\Rod\Rod}^\textrm{s}}
\newcommand{\constMS	}			{\constAttr_{MS}^\textrm{}}
\newcommand{\constMR	}			{\constAttr_{MR}^\textrm{}}
\newcommand{\constBlue}			{\constAttr_{\Rod\Rod}^\textrm{w}}
\newcommand{\constSphere}			{\constAttr_{\Sphere\Sphere}}
\newcommand{\constSR}				{\constAttr_{\Sphere\Rod}}

\newcommand{\funRR}				{\mathcal{F}_{\Rod\Rod}}

\newcommand{\Diameter}			{D}
\newcommand{\Dmin}				{\Diameter_{\Rod}}
\newcommand{\Dmax}				{\Diameter_{\Sphere}}
\newcommand{\dmin}				{d_\mathrm{min}}
\newcommand{\Dav}				{\overline{\Diameter}}

\newcommand{\epsLambda}			{\energy_\lambda}

\newcommand{\FreeEne}				{\mathcal{A}}

\newcommand{\genCoord}			{Q}
\newcommand{\GenCoord}			{\mathbf{\genCoord}}

\newcommand{\Lambdapot}			{\Potential_\lambda}

\newcommand{\Length}				{L}
\newcommand{\Lmin}				{\Length_\Sphere}
\newcommand{\Lmax}				{\Length_\Rod}

\newcommand{\mobility}				{\mu}

\newcommand{\MobilityGen}			{\boldsymbol{\mobility}^\genCoord}

\newcommand{\MobilityLambda}		{\mobility^\lambda}

\newcommand{\nvec}				{n}
\newcommand{\Nvec}				{\hat{\mathbf{\nvec}}}

\newcommand{\orien}				{\hat{u}}
\newcommand{\Orien}				{\mathbf{\orien}}

\newcommand{\pos}					{r}
\newcommand{\Pos}					{\mathbf{\pos}}

\newcommand{\potential}				{\Phi}
\newcommand{\Potential}				{\potential}
\newcommand{\WCA}				{\Potential_\mathrm{rep}}
\newcommand{\PotAttr}				{\Potential_\mathrm{attr}}

\newcommand{\quat}				{q}
\newcommand{\Quat}				{\mathbf{\quat}}

\newcommand{\rand}				{\Theta}

\newcommand{\RandGen}			{\boldsymbol{\rand}}

\newcommand{\Rij}					{\hat{\Pos}_{ij}}

\makeatother
\begin{document}

\preprint{Masson}

\title[Modulators]{
Modulators Selectively Reshape $\alpha$‑Synuclein Phase Transitions}

\author{Holly Masson$^{1,2,3}$}
\author{Massimiliano Paesani$^{1,2,3}$}
\author{Ioana M. Ilie$^{1,2,3*}$}%
 \email[Corresponding author: ]{i.m.ilie@uva.nl}
\affiliation{ 
$^1$ Van 't Hoff Institute for Molecular Sciences, University of Amsterdam, Amsterdam,
 The Netherlands \\ 
$^2$ Amsterdam Center for Multiscale Modeling (ACMM), University of Amsterdam, the Netherlands \\
$^3$ Computational Soft Matter (CSM), University of Amsterdam, the Netherlands
}

\begin{abstract}
Protein phase transitions govern numerous diseases, 
    including neurodegenerative disorders.
In Parkinson’s disease, distinct species of the protein $\alpha$-synuclein 
    undergo phase transitions from disordered to ordered $\beta$-rich states.
The emerging species and transitions between them can be reshaped by modulators, e.g. 
    small molecules, peptides or antibodies. 
Here, we use coarse‑grained simulations to understand the effect of modulators 
    on the thermodynamics and kinetics of $\alpha$‑synuclein phase transitions. 
Each protein is represented as a morphing particle 
    that transforms from a soft sphere (disordered) to a hard spherocylinder ($\beta$‑rich).
Modulators are modeled as soft isotropic particles mimicking peptides. 
Purely repulsive modulators do not 
    alter the final outcome, \ie fibrils form following the same mechanisms independently of modulator concentration.
Attractive interactions towards the disordered protein dose-dependently slow down fibrillization 
    by stabilizing intermediate species, including persistent disordered heteroclusters. 
Specific attraction to the $\beta$-rich state shortens
    fibrils through direct modulator surface "capping" that introduce kinetic barriers to
    monomer templating at the fibril ends and inhibit lateral attachment. 
Overall, these results link modulator properties and environmental conditions to nucleation, 
    fibril elongation and off-pathway trapping, providing a quantitative roadmap for designing interventions that redirect phase transitions toward desirable endpoints.

\end{abstract}

\maketitle

\section{\label{Sec:Intro}Introduction}
Phase transitions, which involve the conversion of a substance from one physical state to another,
    are fundamental phenomena in physical sciences.
In biological contexts, protein phase transitions
    play critical roles in 
    physiological and pathological processes \cite{Ge:advcol_345_103643_2025,Linsenmeier2021,Visser2024,Michaels2023}.
For instance, in neurodegenerative disorders, such as Parkinson's and Alzheimer's, 
    pathologies have been linked to the phase transition process of proteins
    involving $\alpha$-synuclein and amyloid-$\beta$, respectively
    \cite{Zbinden2020,Iadanza2018,Darling2021,Lipiski:sciadv_8_2022,Hardenberg2021,Kffner2021,Linsenmeier2021,Galvagnion2019,Han2024,Rntgen2025,Michaels2023}.
In Parkinson’s disease,  
a hallmark of the disorder is the presence of intracellular protein 
    accumulations known as Lewy inclusions in the brain of diseased patients.
These inclusions 
    and are  composed of distinct species of the protein $\alpha$-synuclein (\asyn)
    \cite{Serratos:mn_59_620_2021,Spillantini:nat_388_839_1997,Hardenberg2021,Raiss2016}.
In the years since its association with Parkinson’s disease 
    was established in the late 1990s, 
    {\asyn} has become a focal point in the study of this disease, particularly the kinetics and thermodynamics associated with pathological phase transitions and the factors that influence them
    \cite{Spillantini:nat_388_839_1997,Polymeropoulos:sci_276_2045_1997,Goedert:natrevneur_2_492_2001,Hardenberg2021}.

{\asyn} is a 140-aminoacid intrinsically disordered protein, 
    which is highly abundant in the brain, 
    localized primarily at presynaptic terminals
    \cite{Maroteaux:jneuroci_8_2804_1988}.
It its native soluble state, 
    {\asyn} lacks a prevalent secondary structure, is highly dynamic
    and can adopt different conformations 
    depending on its environment
    \cite{Maroteaux:jneuroci_8_2804_1988,Weinreb:bchem_35_13709_1996,Jakes:febslett_345_27_1994}.
For instance, it is known to curl into $\alpha$-helical conformations 
    against membranes or detergent micelles, and to adopt 
    $\beta$-sheets when it self-assembles 
    into amyloid fibrils, such as those found in Parkinson’s 
    disease Lewy inclusions
    \cite{Weinreb:bchem_35_13709_1996}.
The central segment of the protein, the non-amyloid component
    is particularly prone to $\beta$-sheet formation, 
    and thus of great interest in the study of {\asyn} phase transitions
    and aggregation
    \cite{Mukherjee:jmb_435_1677132023}.

Aggregation typically proceeds via a nucleation-dependent mechanism 
    involving three kinetic stages: a lag phase, 
    in which critical nuclei form; an elongation phase, 
    during which fibrils rapidly grow via different mechanisms (monomer addition, 
    fibril breakage, secondary nucleation); and a saturation phase, in which monomer depletion leads to a steady-state
    \cite{Wood:jbc_274_19509_1999,Morris:bbapp_1794_375_2009,Shvadchak2015}.
During the lag phase a series of species emerge, \ie 
    soluble (liquid-like) oligomers and protofibrils, some of which evolve 
    into insoluble highly ordered fibrillar structures with a characteristic 
    cross-$\beta$-sheet architecture
    \cite{Weinreb:bchem_35_13709_1996,Ilie:jcp_144_1089_2016,Zhou2024}.
While fibrils are the dominant component of Lewy inclusions, 
    intermediate liquid-like species are increasingly recognized 
    as also being able of disrupting 
    lipid membranes and interfering with intracellular processes
    \cite{Shults:pnas_103_1661_2006,Peelaerts:nat_522_340_2015}.
The formation of fibrillar structures can occur
    following two distinct mechanisms, \ie 
    a homogeneous mechanism, 
    where monomers self-associate (known as primary nucleation), 
    or a heterogeneous mechanism involving a separate surface 
    (for example, a pre-existing fibril or lipid membrane), 
    which lowers the energy barrier for nucleus formation
    \cite{Terakawa:bbamembr_1860_1741_2018,Rawat:bbamembr_1860_18632018}. 
Primary nucleation itself
    can proceed either in one-step (1SN) 
    in which two monomers meet and bind to each other 
    in a single thermodynamic event, or via a two-step (2SN) mechanism, 
    wherein a disordered (often liquid-like) intermediate is involved to catalyze the formation of fibrillar species
    \cite{Ilie:chemrev_119_6956_2019}.
Recent studies indicate that fragmentation of a fibril can also 
    accelerate fibrillization, due to providing new fibril ends 
    where polymerisation can occur
    \cite{Peelaerts:nat_522_340_2015,Morel:bbapp_1867_140264_2019}.
This raises important questions about the optimal approach 
    for intervention during the early stages of Parkinson’s disease. 
The key, it would seem, lies with modulation of the emerging species and transitions between them \cite{Sawner2021,Agarwal2022,Heesink2024,Rodrguez2023}.

Small molecules, peptides and antibodies have shown promise
    as modulators of aggregation \textit{in vitro} and in cellular models
    \cite{Yang:ijbm_230_123194_2023,Oliveri:ejmc_167_10_2019,Singh:fmc_9_1039_2017,Sivanesam:rscadv_5_11577_2015,Perni2017}.
However, the intrinsically disordered nature of $\alpha$-synuclein, 
    and the structural heterogeneity of the emerging species, 
    pose substantial challenges for drug discovery, 
    particularly in identifying stable binding sites or 
    predicting consistent structural responses 
    \cite{Serratos:mn_59_620_2021,Goedert:natrevneur_2_492_2001,Oliveri:ejmc_167_10_2019,Mitra:abb_695_108614_2020}.
Given these challenges in designing modulators experimentally, 
    computational methods have emerged as valuable tools 
    to explore the aggregation pathway
    \cite{Ilie:chemrev_119_6956_2019,Redler:jmcb_6_104_2014,Davidson:jmb_430_3819_2018,SzaaMendyk2023}.
For instance, by integrating atomistic molecular dynamics simulations
    with machine-learning analysis, two promising small molecules 
    able of modulating the interaction between $\alpha$-synuclein and its partner protein 14-3-3$\zeta$ were identified \cite{Chakraborty:ats_2025}.
One, Var84, acts as an orthosteric ligand that likely
    binds at the {\asyn} recognition site and directly competes for binding, while the other, DB11581, functions as an allosteric ligand that influences the interaction indirectly through binding at a remote site on 14-3-3$\zeta$.
Long atomistic molecular dynamics simulations showed that 
    the small molecule fasudil preferentially binds to
    the C-terminal region of $\alpha$-synuclein through a combination of charge–charge and aromatic $\pi$-stacking interactions, while dynamically shuttling among transient binding modes rather than forming multiple stable contacts.
    \cite{Robustelli:jacs_144_2501_2022}.
Rationally redesigned peptides targeting the fibril surface 
    of $\alpha$-synuclein were developed through computational mutagenesis and docking, and later on were shown to effectively suppress 
    fibril and oligomer formation \textit{in vitro} while reducing {\asyn}–induced toxicity in cellular models \cite{Ali:ejmc_289_117452_2025}.
    
While atomistic simulations
    are effective at the monomeric level or for small aggregates, 
    coarse-grained simulations are particularly well suited for gaining
    insight into effects of small agents on {\asyn} phase separation, nucleation, fibrillization or crowding, where the processes occur on long timescales and involve large, structurally diverse assemblies \cite{Ilie:chemrev_119_6956_2019,Tozzini:acr_43_220_2009,Wu:cosb_21_209_2011,Wang2024,SzaaMendyk2023}.
Coarse-grained simulations of $\alpha$-synuclein in crowded environments
    (employing the Martini3 force field) revealed that the protein is highly sensitive to its local environmental context. 
Specifically, in crowded conditions, {\asyn} self-assembly 
    is driven primarily by entropic effects that promote condensate formation, whereas high-salt environments stabilize aggregation through enthalpic interactions that modify chain organization and intermolecular contacts \cite{Wasim:eLife_13_2024}.
One-residue coarse-grained models 
    like CALVADOS \cite{Tesei:calvados_2023} 
    revealed that polyethylene glycol-induced crowding responses depend on sequence charge patterning, highlighting the effects of cellular-like environments on $\alpha$-synuclein compaction and its propensity for phase separation \cite{Rauh:ps_34_2025_2025}.
We developed a single-particle coarse-grained model for intrinsically disordered proteins,
    in which the protein is 
    represented as a morphing particle that switches between two conformational states, \ie a disordered state modeled as a soft isotropic sphere 
    and a $\beta$-rich state modeled as an elongated rod with directional interactions (\Fig{Fig:pot_SR}) \cite{Ilie:jcp_144_1089_2016}.
This model provided mechanistic insight into $\alpha$-synuclein nucleation.
Specifically, that $\alpha$-synuclein nucleation can occur via either 
    direct monomer-to-fibril conversion (one-step) or through metastable intermediates (two-step), with the dominant pathway determined by protein concentration and environmental conditions \cite{Ilie:jcp_144_1089_2016}.
This model was later extended to represent each protein as a chain of 
    morphing particles, enabling a more detailed investigation 
    of $\alpha$-synuclein fibrillar growth \cite{Ilie:jcp_146_115102_2017}.
Specifically, free energy calculations revealed that 
    an attaching protein becomes trapped in a misfolded state, thereby inhibiting
    fibrillar growth until it rearranges to 
    to adopt the fibril-growing conformation, consistent with experimental observations \cite{Wordehoff:jmb_427_1428_2015}.

Here, we build upon the morphing-particle model 
    to investigate how modulators influence $\alpha$-synuclein species formation and phase transitions.
We extend the model by introducing a new type of particle, a small soft sphere, 
    to mimic the interaction of a peptide modulator with {\asyn}. 
By systematically tuning the interactions between the modulator 
    and the disordered and $\beta$-rich proteins, 
    we find that specific and nonspecific interactions selectively reshape the
    emerging
    $\alpha$-synuclein species and their phase transitions. 
Specifically, nonspecific interactions with the disordered protein 
    allow the modulator to act as a bridge between $\alpha$-synuclein monomers, 
    shifting the balance between one-step and two-step nucleation, thereby tuning the lag phase.
This bridging leads to prolonged nucleation timescales 
    and the formation of larger clusters composed predominantly of disordered {\asyn} and modulators.
Specific interactions with the $\beta$-rich proteins influence  
    fibril nucleation and elongation, giving rise to shorter fibrils, 
    whose growth is prevented by the accumulation of modulators on their surfaces.
This protected surface blocks the addition of free peptides and 
    inhibits further fibril elongation.
Together, these distinct modes of modulation, \ie bridging of disordered monomers 
    and selective binding to $\beta$-rich conformers, reshape the kinetics, morphology and stability of $\alpha$-synuclein species and phase transitions.
They provide quantitative insight into what modulator properties can be designed and 
    tuned to control species formation and 
    $\alpha$-synuclein phase transitions.

\section{Computational Methods}
\subsection{{\asyn} and modulator models}	
\begin{figure}[bt]
	\centering
    \includegraphics[width=\columnwidth]{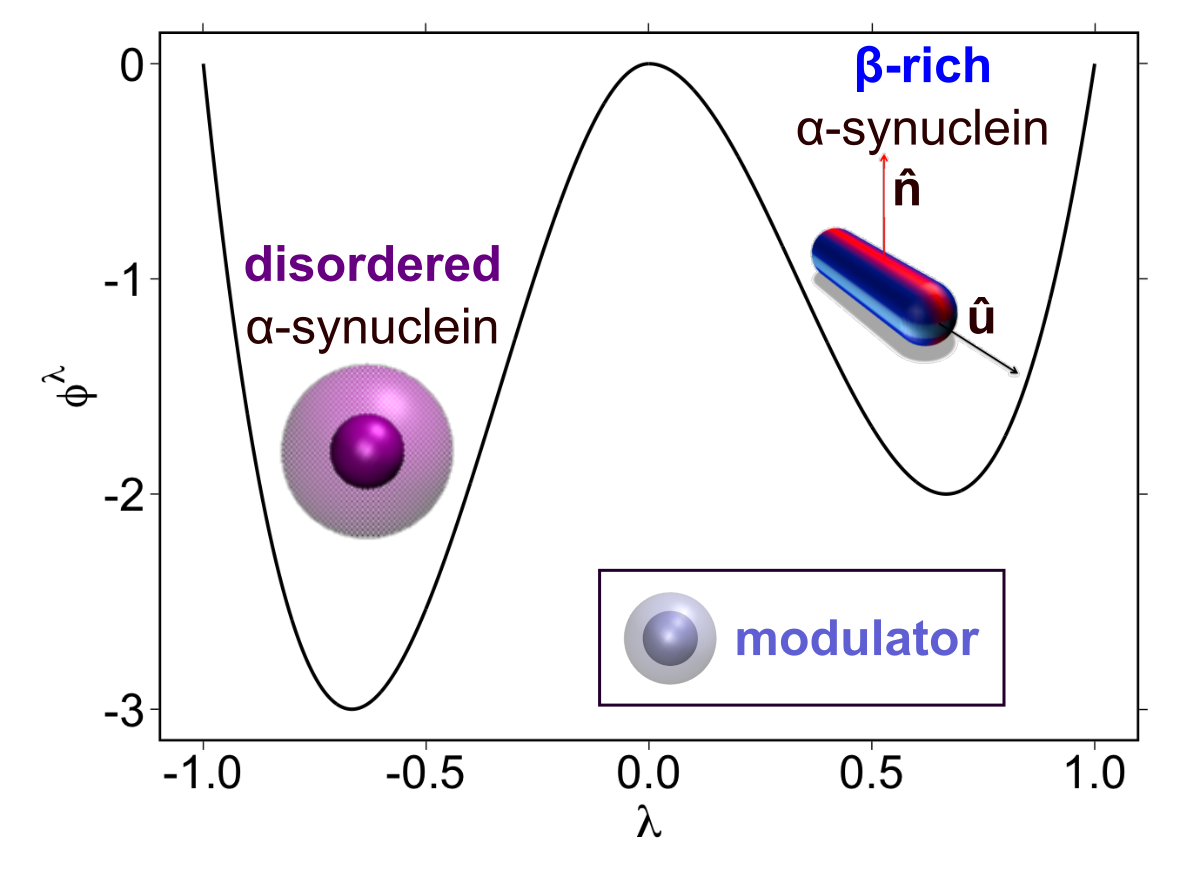}
    \caption{The internal potential plotted against the internal coordinate for
	$\epsLambda^\Sphere =  3$ and $\epsLambda^\Rod =  2$, see \Eq{Eq:LambdaPot}.
	Shown are representative snapshots of {\asyn} in the 
    disordered state (purple sphere) and {\asyn} in the $\beta$-rich ordered state
    (blue spherocylinder with red patches).
    The inset shows the modulator (light blue sphere).
    Note that only the protein particles are morphing between disordered and ordered states according to $\lambda$, 
    while the modulator's state remains unchanged.
	\label{Fig:pot_SR}	}
\end{figure}
We build on a previously introduced model for the
    amyloidogenic core of {\asyn}, which successfully captured
    thermodynamic and kinetic properties of self-assembly,
    including the formation of oligomers and fibrils, as well as nucleation mechanisms \cite{Ilie:jcp_144_1089_2016}.
This model captures the ability of the protein to be 
    intrinsically disordered in solution and adapt to
    attain $\beta$-rich structures in fibrils. 
Hence, in the disordered state, {\asyn} is modeled as a soft sphere 
    with isotropic interactions. 
In the folded state, the protein morphs to become 
    a hard spherocylinder with anisotropic interactions, representing the highly directional 
    contacts associated with $\beta$-sheet stacking (\Fig{Fig:pot_SR}).
Transitions between these states are governed 
    by an internal coordinate $\lambda$ for each particle
    and occur in response to the environment.
In line with experimental findings, 
    the disordered state is favored over the ordered state in solution.
In the simulation, this is accounted for by introducing an internal
    potential, $\Lambdapot(\lambda)$, that favors disordered states 
    for the free particles over ordered states for the $\beta$-rich fibril-bound particles.
Hence, this potential is constructed from two third order polynomials
    that reach local minima of $-\epsLambda^\Sphere$ and $-\epsLambda^\Rod$ for
		$\lambda = \pm \frac23$, and are separated by a local maximum 
        at $\lambda=0$:
\begin{equation}
	\label{Eq:LambdaPot}
		\Lambdapot(\lambda) = 
		\left\{\begin{matrix}
			- 6.75\, \epsLambda^\Sphere \left( \lambda^2 + \lambda^3 \right ) & \textrm{for} \, \lambda < 0 \\ 
			- 6.75 \, \epsLambda^\Rod \left( \lambda^2 - \lambda^3 \right ) & \textrm{for} \, \lambda \geq 0. 
		\end{matrix}\right.
\end{equation}
The relative depth of each well, controlled by the 
    parameters $\epsLambda^\Sphere$ and $\epsLambda^\Rod$ (for sphere and rod, respectively), 
    determines the intrinsic stability of each state 
    in the absence of external interactions. 
As the value of $\lambda_i$ changes, the  
    properties of the particle 
    are able to evolve continuously, endowing the particle with
    the ability to morph between states.
Note that although the folded state is
    unfavourable in solution, it becomes the stable state
	in a fibril.
This stability arises from the attractive interactions
	between proteins. 
In addition, internal states are linked to the geometrical 
    properties of the particles via a smoothed step function
    defined as
\begin{equation}
    \label{Eq:Mu}
    \mu(\lambda) = 
        \left\{
        \begin{array}{ll}
        0, & \lambda < -1, \\[6pt]
        \displaystyle \frac{\tanh(\ALaMu \lambda)}{2 \tanh(\ALaMu)} + \frac12, 
          & -1 \le \lambda \le 1, \\[10pt]
        1, & \lambda > 1
        \end{array}
        \right.
\end{equation}

which ensures that shape changes occur only within a transition region of width $\ALaMu$.
This then translates into the diameter 	
    $\Diameter(\lambda) = \Dmax + \mu(\lambda)(\Dmin-\Dmax)$ and
    length 		
    $\Length(\lambda) = \Lmin + \mu(\lambda)(\Lmax - \Lmin)$ 
    of a particle, with the subscripts $\Sphere$ and $\Rod$
    denoting a sphere and a spherocylinder, respectively.

The current study introduces a third monomorphic ‘modulator’ 
    particle to the model, capable of interacting with the protein.
In contrast to the polymorphic {\asyn} particle, 
    which can dynamically transition 
    between disordered and ordered states 
    via an internal coordinate, 
    the modulator particle introduced in this study 
    is modelled as a single coarse-grained soft sphere, 
    with a fixed diameter and fixed length.

\subsection{Interaction potentials}
Due to the polymorphism of the particles, the interaction 
    between different {\asyn}-proteins
    is also described by an internal state-dependent potential.
This is particularly important in a system such as this, 
    where particles morph from soft isotropic spheres 
    (representing a disordered protein)
    to hard anisotropic spherocylinders 
    (representing a ordered protein), 
    as the type and strength of steric interactions differ significantly between these two states.
Since the particle’s conformation, and thus 
    its effective shape and properties, vary continuously as a function of the internal coordinate $\lambda_i$, the interaction dynamically 
    adapts to these changes. 

The repulsion between any two particles $i$ and $j$ 
    characterized by their center of mass positions, $\Pos_i$, $\Pos_j$, orientations,
    $\Orien_i$, $\Orien_j$, and internal states,
    $\lambda_i$,$\lambda_j$,
    is computed
    based on the minimum distance between their long axes 
    $\dmin(\Pos_i, \Pos_j, \Orien_i, \Orien_j,\lambda_i,\lambda_j) = \min_{\alpha_i, \alpha_j}
		 \left | \left ( \Pos_i + \alpha_i \Orien_i \right ) - 
				\left( \Pos_j + \alpha_j \Orien_j \right ) \right |$ 
    \cite{Vega:cc_18_55_1993,Ilie:jcp_144_1089_2016,Ilie:jcp_146_115102_2017}.
Here, $\Orien_i$ and $\Orien_j$ are unit vectors 
		pointing along the long axes of the particles and
        $\alpha_i $ and $\alpha_j$ are two minimization parameters obtained from  
		$|\alpha_i| \le \frac12(\Length_i-\Diameter_i)$ and
		$|\alpha_j| \le \frac12(\Length_j-\Diameter_j)$, respectively.

The repulsive potential accounts for the 
    softness of the particles, which depends on their
    internal state. 
As such, the repulsion between spherical particles 
    is softer than between spherocylindrical ones, reflecting the looser packing of the protein in its disordered state compared to the more compact ordered state.
Furthermore, the repulsive potential should become significant,
		{\ie} increase above the thermal energy $\kBT$
			with $\kB$ Boltzmann's constant and $T$ the temperature,
		when the minimum distance between two particles 
        is shorter than 
		the average of their radii, $\Dav = \frac12 ( \Diameter_i + \Diameter_j )$. 
The repulsive potential between any two particles then reads
		\begin{equation}
	\label{Eq:newWCA}
		\frac{\WCA^\lambda}{\kBT}=
		\left\{\begin{matrix}
			 \displaystyle
 			 \left ( \frac{2\Dav}{\dmin} -1 \right )^{n}  & \textrm{for} \,\,\,\dmin < 2\Dav \\
			0			& \textrm{for} \,\,\, \dmin > 2 \Dav,
		\end{matrix}\right.
	\end{equation}
	where the exponent $n = 4 + ( \mu_i + \mu_i )$, \ie the hardness of the potential,
    was selected to vary from 4 to 6
		for {\asyn}-{\asyn} interactions.
For {\asyn}-modulator repulsion the exponent was set to 5 
    to reflect the structural compactness of a modulator, which 
    is greater than that of a disordered protein, 
    yet less tightly arranged than the folded 
    conformation of {\asyn}.

The accumulation of the proteins is driven by 
    attractive potentials, designed to reflect
    van der Waals, hydrophobic interactions 
    and hydrogen bonding between the proteins. 
Importantly, the strength and character of these interactions 
    depend on the distance and orientations
    between particles, 
    and also on their conformational states. 
Disordered {\asyn}, owing to their larger radii of gyration and
    packing, 
    interact via weaker, longer-ranged and mostly isotropic forces. 
In contrast, folded $\beta$-rich proteins form compact, 
    anisotropic structures where short-ranged, 
    directional hydrogen bonding dominates, consistent with 
    $\beta$-sheet stacking in fibrils. 
The modulators are modeled as isotropic particles, 
    mimicking the structural compactness of a 
    folded polypeptide domains.
The proposed potential $\PotAttr = \PotAttr^\lambda + \PotAttr^M$ 
    encompasses the interactions between
    between {\asyn} proteins, $\PotAttr^\lambda$, and interactions between
    the {\asyn} molecules and the modulator, $\PotAttr^M$.

The {\asyn}-{\asyn} potential, $\PotAttr^\lambda$, is constructed 
    as the sum of three contributions \cite{Ilie:jcp_144_1089_2016,Ilie:jcp_146_115102_2017}, 
	\begin{align}
		\label{Eq:Attr}
		\begin{split}
		\frac{\PotAttr^\lambda}{\kBT}
			= \eta(\dmin)
			& \Big\{  
			\constSphere \left ( 1-\mu_i \right ) \left( 1-\mu_j \right)  \\
			&  +
			 \funRR \mu_i \mu_j \\
			&  + 
			\constSR \Big[\mu_i (1-\mu_j ) + \mu_j (1-\mu_i) \Big] \Big\},
		\end{split}
	\end{align}	
    each associated with a different type of interaction: disordered-disordered (sphere-sphere), governed by
    $\constSphere$, 
    ordered-ordered (rod-rod), governed by $\funRR$, 
    and disordered-folded (sphere-rod), defined through $\constSR$, pairs. 
This potential is scaled by a distance-dependent 
    function,  
    $\eta(\dmin) = \frac{\tanh(\AAttr (\dmin - 2\Dav))}{\tanh(2\AAttr \Dav)} $, which ensures
    that the attraction 
    turns off smoothly at the cut-off range
    $\dmin < 2\Dav$.
The steepness of the transition between $\eta(0)=-1$ 
    and $\eta(2\Dav)=0$ is given by the parameter
    $\AAttr = \aAttr (\mu_i+\mu_j) + \bAttr$,
			    with $\aAttr > 0$ and $\bAttr > 0$.
The potential in \Eq{Eq:Attr} reduces to a purely distance
    dependent attraction with a minimum at $-\constSphere$
    for two spheres with $\mu_i \simeq  \mu_j \simeq 0$.

The interaction between two rods with $\mu_i \simeq  \mu_j \simeq 1$ is described by the second term on the \rhs, 
    with $\funRR = \left[ \constBlue + g_{ij}  \constRed \right]$ accounting also for the relative
    orientations of the ordered proteins.
Here $f_{ij}$ and $g_{ij}$ are two directional functions that
    describe the parallel arrangement of the rods and their
    hydrogen bonding ability, respectively.
The two constants $\constBlue$ and $\constRed$ 
    represent the weak van der Waals and hydrophobic interactions, and the strong hydrogen bonds between two proteins, resppectively. 
Essentially, $f_{ij} \in [0,1]$ is an alignment factor
    that accounts for the parallel alignment of two rods
    $(\Orien_i \cdot \Orien_j)^2$ and the sliding of the rods
    along their long axes $ h(\Orien_i \cdot \Rij)	h(\Orien_j \cdot \Rij)$ and is thus defined as
    $f_{ij} = (\Orien_i \cdot \Orien_j)^2
			h(\Orien_i \cdot \Rij)
			h(\Orien_j \cdot \Rij)$.
Here $h(x) = 1-ax^2$ for $x^2 < 1/a$ and 0 otherwise
    captures the variations in energy 
    as one rod is displaced
		along its long axis, at fixed orientation and $\dmin$.
The interaction is the most favorable
    when the connecting vector $\Pos_{ij}$ is perpendicular to both rods.
The directional function $g_{ij} \in [0,1]$  
    accounts for the hydrogen bonding ability 
    between two proteins.
This is achieved by endowing the particle with 
    an orientation vector, $\Nvec$, perpendicular to the long axis $\Orien_i$, and favoring its alignment 
    in parallel to the
	center-to-center direction $\Rij$, resulting in
    $g_{ij}= (\Rij \cdot \Nvec_i)^p (\Rij \cdot \Nvec_j)^p$, with
    $p$ an integer. 

The attractive potential, $\PotAttr^M$,
    between modulator particles and {\asyn} particles 
    is formulated in a simplified manner to represent isotropic interactions with both disordered and ordered {\asyn}:
\begin{align}
		\label{Eq:AttrM}
		\begin{split}
		\frac{\PotAttr^M}{\kBT}
			= \theta(i,j) \eta(\dmin)
			& \Big\{  
			\constMS \left ( 1-\mu_i \right ) \left( 1-\mu_j \right) \\
            + & 
            \constMR \left[ \mu_i \left ( 1-\mu_j \right )  + \mu_j \left( 1-\mu_i \right) \right]
            \Big\},
		\end{split}
	\end{align}	
with $\theta(i,j)$ a step function that returns 1 if
    any of the two particles is a modulator and 0 otherwise.
Modulators are modelled as uniform spheres,
    which do not change properties, 
    and thus their internal state is fixed in the simulations to 
    $\mu = 0$.
Therefore, the attraction to the {\asyn} molecules
    is described by a distance dependent potential,
    which captures the binding abilities to
    disordered and ordered proteins. 
Similarly as for two {\asyn} proteins, the distances
    dependent function $\eta(\dmin)$ will scale with the
    properties of the interacting protein.
Since both terms on the \rhs of \Eq{Eq:AttrM} 
    scale with the average diameter of the interacting particles, 
    the interaction between a modulator and a spherical particle, which has a minimum at $\constMS$,
    is longer ranged than that between a modulator and an ordered particle, with a minimum at $\constMR$.
    
\section{Brownian Dynamics Simulations}
All simulations were carried out using Brownian dynamics
    to simulate the translational, rotational and 
    internal dynamics of the particles.
Here, we limit ourselves at presenting just the 
    generalized equation of motion as the 
    three independent equations have been previously
    introduced \cite{Ilie:jcp_142_114103_2015,Ilie:jcp_144_1089_2016,Ilie:jcp_146_115102_2017}.
The generalized equation of motion is given by \cite{Oettinger,Gardiner}
	\begin{align}
	\label{Eq:GEoM}
		\GenCoord(t+\dt) - \GenCoord(t) 
		 = 
		&- \MobilityGen \frac{\partial \FreeEne}{\partial \GenCoord}\dt \nonumber \\
		&+ \kBT \frac{\partial}{\partial \GenCoord} \cdot \MobilityGen \dt \nonumber \\
		&+ \left( \MobilityGen \right )^{1/2} \RandGen^\genCoord(t) \sqrt{2\kBT \dt},
	\end{align}
	where $\GenCoord$ represents the full set of generalised coordinates.
The first term on the {\rhs}
		describes the displacements of $\GenCoord$
        over a time-step $\dt$, arising from the balance between
		the thermodynamic force $\boldsymbol{\mathcal{F}} = -\partial \FreeEne/\partial \GenCoord$,
		where $\FreeEne$ is the free energy as a function of $\GenCoord$, and
		the opposing solvent friction, which is the inverse of the mobility matrix $\MobilityGen$.
The second term accounts for inhomogeneity of the 
    mobility tensor $\MobilityGen$, which leads to a non-zero contribution even in the absence of a deterministic force.
The last term corresponds to Brownian displacements of the
    generalized coordinate.
The components of the time-dependent Markovian vector 
    $\RandGen^\genCoord$
	are uncorrelated, have zero mean, unit variance.
The size of these random displacements 
    is connected to the mobility tensor via
	the fluctuation-dissipation theorem.
The three decoupled equations of motion can be readily
    derived by replacing the generalized coordinate
    with the positions of the center of mass of 
    the particles $\Pos$, 
    their orientations $\Quat$ and internal coordinates $\lambda$ for translation, rotation \cite{Ilie:jcp_142_114103_2015}
    and internal states \cite{Ilie:jcp_144_1089_2016,Ilie:jcp_146_115102_2017}, respectively.   
	
\section{Simulation details}
The properties of the particles were selected 
    as in the original model to
    match experimental values as much as possible
    \cite{Ilie:jcp_144_1089_2016}.
We focus on the central 30-90 segment of
    {\asyn}, which is key in driving phase
    transitions and fibrillization.
Given that a protein in its disordered conformation 
    is slightly more compact than a random coil of similar length, yet less compact than a tightly folded structure, we approximate the radius of gyration for the 30–90 segment as 2.61 nm. 
In the folded conformation, we approximate the length
    of a spherocylinder $L_R=3.5$~nm
    as the average length of the five 
    $\beta$-strands, inspired from solution NMR studies
    \cite{Vilar:pnas_105_8637_2008}.
The diameter of a rod $D_R=2.5$~nm
    was set as the average between the combined width of the five $\beta$-strands ($\approx$ 4 nm) and the distance ($\approx$ 1 nm) between 
    two proteins along the fibril axis \cite{Ilie:jcp_144_1089_2016,Ilie:jcp_146_115102_2017}.
To ensure that the transition between two different states is 
    limited to a small window, we use $\ALaMu$ = 10. 
The depth of the double well potential was fixed at $\epsLambda^\Sphere=3$ and $\epsLambda^\Rod=2$ for a disordered and an ordered protein, respectively.
This choice is motivated by the fact that previous simulations
    have shown that these parameters are efficient in capturing
    different nucleation pathways as well as fibrillar growth \cite{Ilie:jcp_144_1089_2016,Ilie:jcp_146_115102_2017}.
The radius of gyration of the modulator ($\approx$ 1.43~nm) was 
    chosen \textcolor{black}{to match that of a typical folded 
    polypeptide shorter in sequence than {\asyn}, \ie roughly half the radius of gyration of the disordered {\asyn} particle.}
    
For the {\asyn-\asyn} interaction potentials, the parameters 
    were selected to
    match the original model \cite{Ilie:jcp_144_1089_2016}.
As such, the steepness of the attractive potential was fixed
    at $\AAttr= 0.2 + 0.4*(\mu_i + \mu_j)$, with Aattr in
    units of nm$^{-1}$.
The parameters in \Eq{Eq:Attr} were chosen such that
    the relevant mechanisms in absence of any modulators are preserved, $\constSphere=3$, $\constSR=0$, $\constBlue=0$ and $\constRed$=20.
To understand the effects of the modulators, we 
    varied the concentrations of added modulator, $\cMod$,
    (0 to 37.5 $\mu$M) as well as 
    their affinities towards the disordered and the ordered
    {\asyn}, $\constMS$ and $\constMR$, respectively.
    
Simulation systems were initialized by placing 200 
    disordered ($\lambda = - 2 / 3$) {\asyn} particles, with random positions and orientations and
    a variable number of modulators
    in a cubic simulation box with periodic boundary conditions.
All simulations were carried out in implicit solvent, at 
    the water viscosity of $\eta = 7 \cdot 10^{-4}$\,Pa\,s,
    a temperature of 310\,K and using a timestep $\dt=1.3\cdot 10^{-13}$s for 52 $\mu$s per system.
The mobility of the particles were calculated  on the fly
    \cite{Ilie:jcp_144_1089_2016}
    and the internal mobility of the {\asyn} particles
    was fixed at $\MobilityLambda=1.4 \cdot 10^{30}$~J/s,
    which ensures that transitions are rare events yet numerous transitions occur during a simulation. 

\section{Results and Discussion}
\subsection{Modulator-independent phase transitions}
\begin{figure}
\includegraphics[width=\columnwidth]{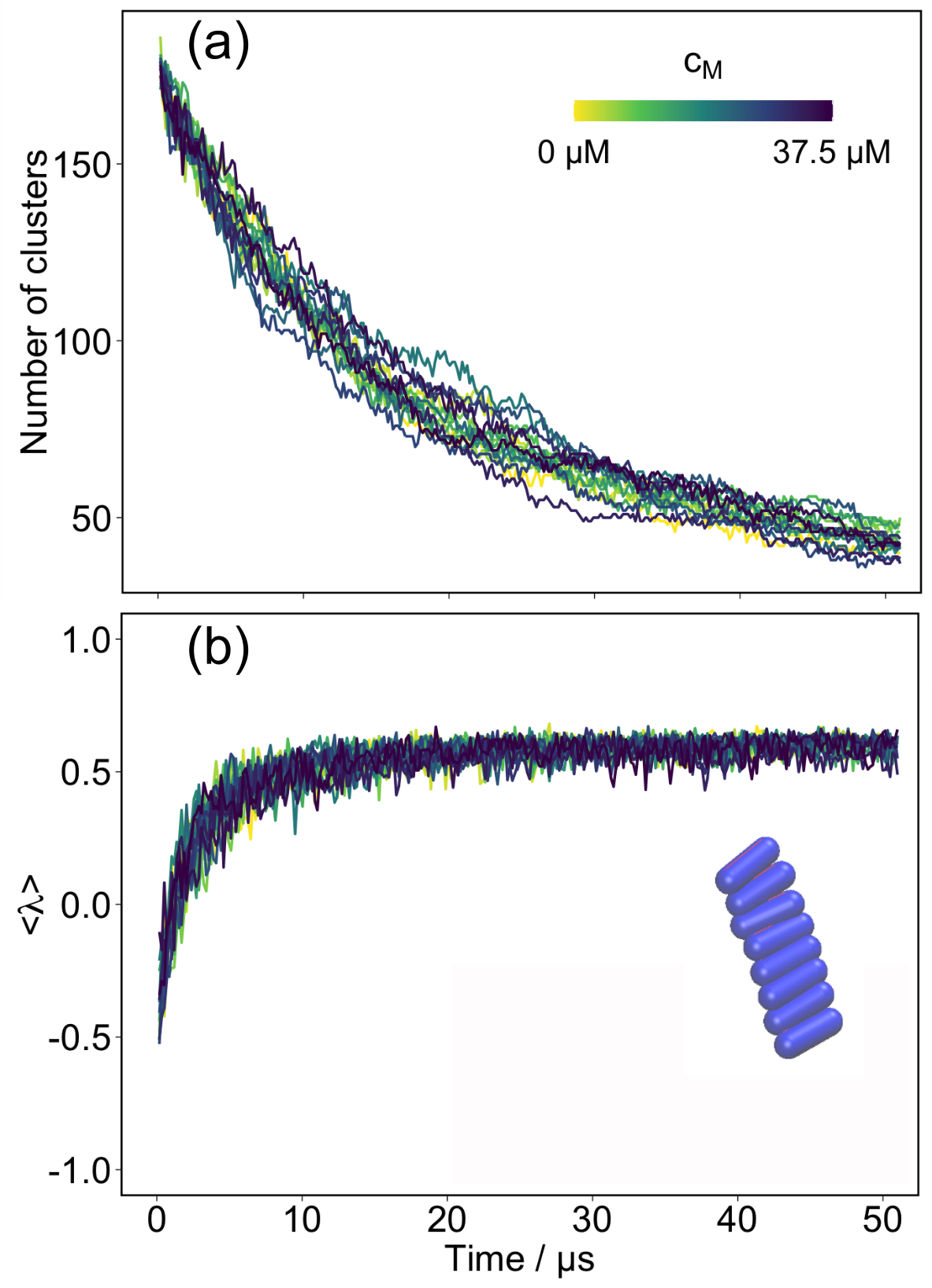}
\caption{\textbf{Effects of repulsive modulators on {\asyn} clustering.}
    (a) Number of clusters as a function of time at different modulator concentrations.
    (b) Time series of the average protein conformation in the simulation box at different modulator concentrations.
    \label{Fig:rep}}
\end{figure}
To assess the effects of modulator concentration 
    on $\alpha$-synuclein assembly, 200 disordered proteins were placed in a cubic simulation box at a concentration of 50 $\mu$M
    and the modulator concentration, $\cMod$, was systematically varied between 0 $\mu$M and 37.5 $\mu$M.
In first instance, only repulsive interactions 
    between modulators and the protein were included, mimicking the
    effect of crowders in the semi-dilute regime \cite{Bhattacharya:bpjrev_5_99_2013}.
The analysis focused on the temporal evolution of the number of clusters
    revealed a rapid decrease in the number of emerging clusters within the first few
    $\mu$s, which then levels toward a comparable number of clusters, 
    independently of the modulator concentration (\Fig{Fig:rep}(a)).
The profiles follow similar exponential decays aligning 
    along the same averaged profile, and reach comparable cluster numbers (ca. 50), indicating
    that the modulator concentration
    has marginal impact on the aggregation kinetics.
Simultaneously, the average internal conformation of the proteins, 
    $<\lambda>=\frac{1}{N}\sum_{i=1,N} \lambda_i$ over particles 
    $i=1..N$,
    shows an increase from $\approx$ -0.5 to $\approx$ +0.5, corresponding
    to overall transitions from disordered to ordered states, respectively (\Fig{Fig:rep}(b)).
In agreement with the cluster analysis, 
    visual inspection indicates 
    fibrillar structure formation across all systems, independent of repulsive modulator concentration (snapshot in \Fig{Fig:rep}(b)).
\begin{figure*}
\includegraphics[width=2\columnwidth]{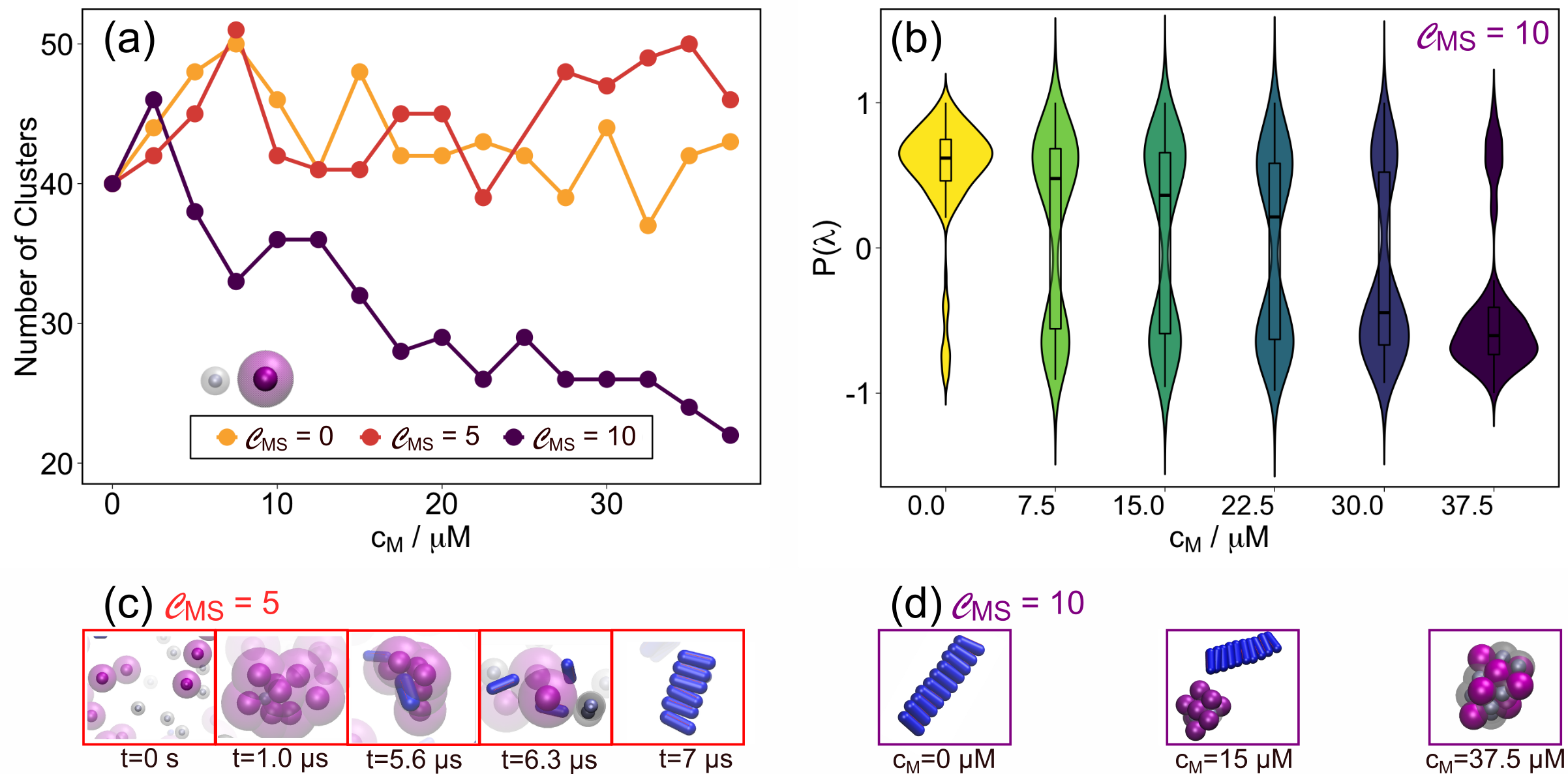}
\caption{\textbf{Effects of modulator concentration and non-specific binding to the disordered protein.}
    (a) Number of emerging clusters as a function of modulator concentration.
    (b) Violin plot of the protein conformations at high affinity of the modulators towards the disordered proteins. 
    (c) Snapshots highlighting two-step nucleation at moderate modulator-protein attraction, $\constMS$=5 and $\cMod$=37.5 $\mu$M.
    (d) Representative snapshots of the emerging structures at $\cMod$=0 $\mu$M, $\cMod$=15 $\mu$M and $\cMod$=37.5 $\mu$M.
    \label{Fig:attrS}}
\end{figure*}
Additionally, we find that fibril nucleation can occur both in
    one step as well as in two steps.
More specific, multiple ordered proteins come 
    into proximity and spontaneously assemble into a fibril nucleus (one-step nucleation), or nucleation can proceed via oligomer formation, followed by the conversion into growth‑competent nuclei through structural rearrangement (two-step nucleation).
While at low modulator concentrations, nucleation appears 
    largely unchanged, at higher concentrations, visual inspection shows that the modulators can promote oligomer formation.
The emerging oligomers either disintegrate, convert into fibrils or grow into larger 
    assemblies. 
We ascribe this to the fact that modulators can 
    effectively increase the local protein concentration, raising the 
    probability of forming metastable clusters.
Here, locally increased modulator concentrations can stabilize small oligomers 
    just enough to extend their lifetimes before conversion or dissociation.

\subsection{Nonspecific attraction tunes conformational ensembles and emerging species}
Experiments have shown that non-specific 
    agents can modulate
    $\alpha$‑synuclein dynamics and self-assembly \cite{Heravi:ps_33_2024,Lipiski:sciadv_8_2022}.
To investigate the effects of such agents, 
    we endowed the modulators with isotropic attractions.
To achieve this, we systematically varied the depth of the attractive well by modifying
    $\constMS$ from 0 to 10, thereby 
    describing weak to strong binding towards the disordered protein.
The analysis focused on the emerging protein cluster size indicates that, under purely repulsive modulation, the system yields ca. 40–50 aggregates independently of the modulator concentration (\Figs{Fig:attrS}(a), \textcolor{black}{S1(a)}, yellow profile).
A moderate attraction has marginal impact on the 
    global cluster-size statistics, yielding a comparable number of clusters as in the case of repulsive modulators (\Figs{Fig:attrS}(a), \textcolor{black}{S1(a)}, orange profile).
\textcolor{black}{
The comparable number of emerging clusters is consistent with the strength of the pair interaction
relative to the concentrations sampled. 
For $\constMS = 5$, we estimate an effective pairwise modulator-{\asyn} association constant
    $K_a = \int \left[ e^{-\left(\PotAttr^M(d) + \WCA(d)\right)/\kBT} - 1 \right] dV$,
    yielding a dissociation constant $K_d = 1/(N_A K_a) \approx 137~\mu$M 
    (with $N_A$ the Avogadro constant), which is almost four times the 
    highest modulator concentration used here ($\cMod = 37.5~\mu$M). 
Modulator binding is therefore expected to remain dynamic in this regime, 
    consistent with the frequent attachment and detachment events 
    involving individual proteins observed in the trajectories.
At higher attraction strengths,
    for which $K_d$
  falls inside the range of concentrations sampled ($\constMS\ge$8),
  leads to a concentration-dependent 
    decrease in number of emerging clusters (Fig. S1(a)).}
Specifically, increasing the modulator concentration leads to 
    a decrease in the number of clusters from $\approx$ 40 to $\approx$ 20 
    for 0 $\mu$M to 37.5 $\mu$M, respectively (\Fig{Fig:attrS}(a), purple profile).

\textcolor{black}{At moderate attraction strength, the modulators do not substantially alter the 
    final aggregation outcome as the emerging protein clusters remain predominantly fibrillar
    (\textcolor{black}{Fig. S2}).
To determine whether the modulators nevertheless alter the 
    nucleation pathway relative to the modulator-free system, 
    we zoomed-in on the two-step nucleation mechanism at 
    intermediate modulator attraction and $\cMod = 37.5~\mu$M (\Fig{Fig:attrS}(c)).}
Within the first microsecond, multiple 
    disordered proteins coalesce into a dynamic cluster.
Subsequently, over the next 4 $\mu$s, 
    the fluid-like assembly remains largely intact.
A subset of proteins reorganize and adopt
    $\beta$-rich conformations, which then engage in cooperative association to result a fibrillar structure.
Compared with previous results under identical conditions
    \cite{Ilie:jcp_144_1089_2016} in absence of modulators, the nucleation kinetics is slower, which increases the duration of the lag phase.

\begin{figure*}
\includegraphics[width=2\columnwidth]{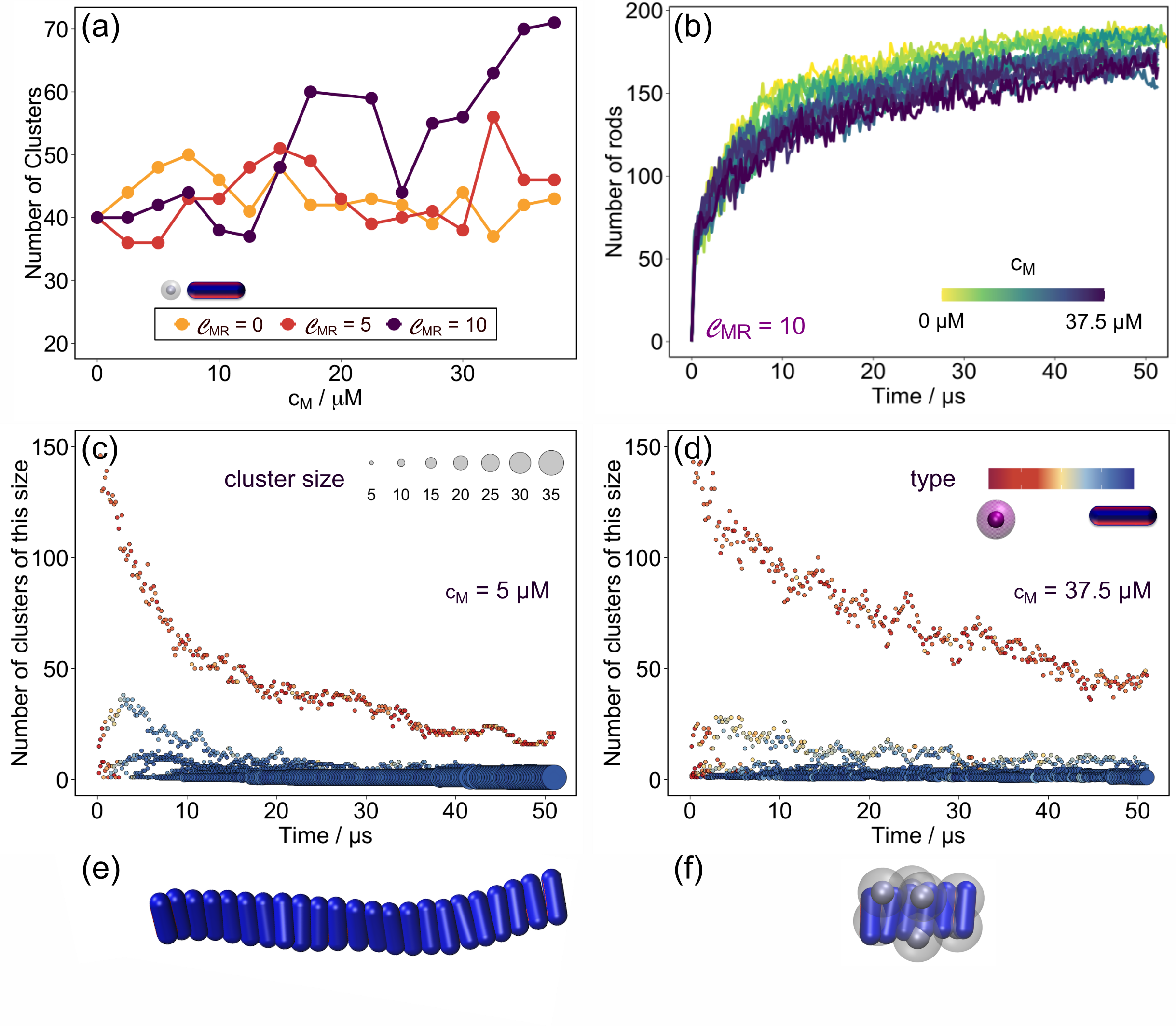}
\caption{\textbf{Effects of $\beta$-specific modulators on {\asyn} fibrillization.}
    (a) Number of protein clusters as a function of modulator concentration.
    (b) Time series of the number of $\beta$-rich proteins at different modulator concentrations and high modulator-$\beta$-rich protein affinity, $\constMR$=10.
    (c-d) Time series of the type and size of the emerging clusters at $\cMod$=5$\mu$M and $\cMod$=37.5$\mu$M, respectively.
    The symbol color encodes cluster type, \ie red for disordered, blue for $\beta$-rich and yellow for intermediate internal states,
    and symbol area encodes the cluster size.
    (e) Representative snapshot of an $\alpha$-synuclein fibril at $\cMod$=5$\mu$M and
    (f) at $\cMod$=37.5$\mu$M. At high modulator concentration the emerging fibrils are shorter due to the accumulation 
    of modulators on the fibril surface thereby blocking incoming monomers to attach and continue fibrillar growth.
    \label{Fig:attrR}}
\end{figure*}    

To investigate the structure of the emerging clusters, 
    we analyzed the protein conformations relying
    on the internal parameter $\lambda$.
The results show that
    at modulator concentrations below 7.5 $\mu$M, the system
    is mainly populated by $\beta$-rich proteins ($\lambda>0$),
    which indicates that the formed clusters correspond primarily to
    fibrillar structures (\Fig{Fig:attrS}(b)).
As modulator concentration is increased ($\leq$ 30 $\mu$M), 
    the $\lambda$-distribution gradually
    shifts towards disordered species as well ($\lambda<0$), indicating a 
    coexistence of disordered and $\beta$-rich structures.
At high modulator concentrations (37.5 $\mu$M), 
    the $\lambda$-distribution shows that
    the disordered state is favoured ($\lambda<0$).
In combination with the corresponding number of clusters, this indicates the
    emergence of predominantly oligomeric species.
Visual inspection of the simulations reveals 
    that disordered proteins heteroassemble with modulators, 
    which help stabilize the assembly (Movie S1, snapshots in \Fig{Fig:attrS}(d)). 
This interaction promotes the progressive growth of the oligomer through monomer addition.
The stabilized oligomer grows predominantly by
    stepwise monomer addition, rapidly exceeding the critical size of six monomers for conversion into a fibrillar nucleus \cite{Ilie:jcp_144_1089_2016}, which inhibits the two-step nucleation mechanism.
    
All in all, our results indicate that
    nonspecific attraction towards the disordered state, equivalent to hydrophobic interactions, drives a concentration-dependent shift in the distribution of assembled species and modulates the kinetics of fibril nucleation.

\subsection{$\beta$-selective modulators inhibit fibril elongation through surface "capping"}
A therapeutically motivated strategy is 
    to design binders that recognize $\beta$–rich conformations 
    of $\alpha$‑synuclein and thereby modulate fibril growth and seeding \cite{Sangwan:elife_9_2020,Hjelm:front_16_2025,Ali:ejmc_289_117452_2025_2025}.
To understand the underlying mechanisms of such agents, we 
    systematically varied the modulator affinity towards the $\beta$–rich conformations and investigated the effects on fibril nucleation and elongation.

The analysis focused on the number of clusters as a function of modulator concentration
    concentration, shows that at low to medium
    modulator-$\beta$-rich {\asyn} interactions,
    both the cluster count and their average internal state remain largely unchanged
    with increasing $\cMod$, while
    the emerging species are predominantly $\beta$-rich (\Fig{Fig:attrR}(a), \textcolor{black}{Fig. S1 (b)}).
At high affinity,
    the number of emerging clusters nearly doubles (from 40 to 70)
    in a dose-dependent manner at modulator concentrations over
    12.5 $\mu$M (\Fig{Fig:attrR}(a), purple line).
Although the final products are primarily fibrillar independently of $\cMod$ 
    (\textcolor{black}{Fig. S3} (a), (c)), the kinetics of the process
    is affected by the modulator concentration (\textcolor{black}{Fig. S3 (b), (d)}).
Specifically, tracking the time evolution 
    of $\beta$-rich proteins reveals a rapid increase in the total number of
    rods within the first 10 $\mu$s, reflecting the fast conversion of disordered
    proteins into ordered species, ascribed to fibril nucleation (\Fig{Fig:attrR}(b)).
This growth then levels off into a plateau, consistent with the establishment 
    of a steady population of $\beta$‑rich proteins. 
Comparing systems without modulators to those with modulators 
    reveals that, while the onset of nucleation is marginally affected, 
    the subsequent increase in rod number proceeds slower with increasing $\cMod$ (\Fig{Fig:attrR}(b)).

Zooming in on the type and sizes of the emerging clusters shows that 
    all systems transition from many small disordered clusters at early times (red symbols in \Fig{Fig:attrR}(c-d)) to a few larger $\beta$‑rich clusters at late times (blue symbols in \Fig{Fig:attrR}(c-d)).
At low modulator concentration,
    disordered proteins accumulate into clusters up to ten monomers 
    within the first $\approx$ 5 $\mu$s (red symbols in \Fig{Fig:attrR}(c)).
A fraction converts directly to $\beta$‑rich fibrils 
    (red $\rightarrow$ blue), while
    other clusters first form mixed intermediates 
    (red $\rightarrow$ yellow)
    and undergo
    structural rearrangements
    before maturing into $\beta$‑rich structures (yellow $\rightarrow$ blue).
As the modulator concentration is increased, 
    the size of the emerging clusters decreases (smaller blue symbol size in \Fig{Fig:attrR}(d) as compared to the larger blue symbols in \Fig{Fig:attrR}(c)).
Specifically, at $\cMod$=37.5$\mu$M in the first 10 $\mu$s small
    clusters consisting of both $\beta$-rich and disordered proteins form (\Fig{Fig:attrR}(d)).
Many of these evolve into fibrillar structures with less than 20 monomers in their
    composition, which persist until the end of the simulation.
Relative to systems with fewer modulators, this corresponds to a significant 
    reduction  fibril length, \ie at $\cMod$=5$\mu$M, fibrils reach lengths of about 35 proteins (\Fig{Fig:attrR}(c)), whereas at $\cMod$=37.5$\mu$M they extend only up to roughly 20 proteins.
Interestingly, modulator-coated fibril surfaces can occasionally bridge adjacent 
    fibrils, facilitating lateral association and bundle formation.
Additionally, the population of free, uncomplexed disordered proteins 
    is higher at high modulator concentration (red symbols in \Fig{Fig:attrR}(d-e)).
    
All in all, our results show that the {\asyn} proteins, which
    convert into $\beta$‑rich conformations
    present high affinity surfaces for the modulators.
These proteins can associate to form fibrils, which themselves act as attractive
    scaffolds for the modulators.
As such the modulators  
    can accumulate on the fibril surface.
At sufficiently high modulator concentrations, the adsorbed
    modulators act as surface “cappers” by forming a protective coating 
    that can
    sterically or energetically prevent further incorporation 
    of monomers, thereby inhibiting fibril elongation
    (\textcolor{black}{Movie S2}).
This hypothesis is further supported by the increased population of free, 
    disordered proteins at high modulator concentration, which indicates that monomers remain in solution rather than being sequestered into growing fibrils.


\section{Discussion}
More than 50 million people worldwide are affected by 
    neurodegenerative disorders, such as Parkinson's disease.
The underlying disease mechanisms are linked to  
    phase transitions of the 
    protein $\alpha$-synuclein, an intrinsically disordered protein also known as \textit{protein chameleon} \cite{Uversky:jbc_14_10737_2001}.
$\alpha$-synuclein can undergo phase transitions 
     giving rise to a wide spectrum of emergent species, ranging from liquid-like condensates and oligomers to $\beta$‑rich fibrils.
To understand the emergence of these species and the effects of molecular 
    modulators on their formation mechanisms and kinetics, we performed 
    coarse-grained
    simulations of $\alpha$‑synuclein \cite{Ilie:jcp_144_1089_2016,Ilie:jcp_146_115102_2017} 
    in the presence of modulators. 
In this model, the protein is represented as a single, 
    polymorphic particle that switches between a soft, isotropically interacting sphere representing the disordered state and an elongated spherocylinder with directional, hydrogen‑bond–like attractions representing the $\beta$‑rich state.
The modulator is modeled as a single soft sphere with isotropic interactions representing
    a peptide
    \textcolor{black}{about half the size of the protein}.
Throughout our simulations, we systematically varied 
    modulator concentration (0--37.5 $\mu$M) along with its relative affinity (0--10 $\kBT$) towards disordered versus ordered, $\beta$‑rich states.
Our results show that modulator concentration and interactions 
    reshape the 
    kinetic and thermodynamic pathways of {\asyn} self-assembly, 
    fibril nucleation and elongation.
The results are three-fold.
    
First, soft repulsive modulators act as 
    as low‑dose crowders that do not alter the formation of fibrils nor change the underlying mechanisms.
Hence, fibril nucleation occurs both in one‑step 
    and in two‑steps via dynamic disordered intermediates, and both mechanisms persist across different modulator concentrations.
Our results are consistent with experiments showing the formation 
    of fibrillar species also when in solution with different
    crowding agents (proteins, polysaccharides and polyethylene glycols) \cite{Uversky:febslet_515_1873_2002,Horvath:bpj_120_3374_2021}.
While these experiments show a strong acceleration of 
    fibrillation in presence of crowding agents, our simulations show marginal influence on the aggregation mechanisms when increasing the concentration to 37.5 $\mu$M. 
This is most likely because we are in the semi-dilute regime,
    significantly below concentrations commonly used experimentally, where crowding is at least tenfold higher
    \cite{Uversky:febslet_515_1873_2002,Horvath:bpj_120_3374_2021}.
\textcolor{black}{
Additionally, we expect modulator size to provide an important additional 
    control parameter for the observed mechanisms \cite{Latshaw2014}.
In the present model, the modulator diameter is approximately half that of the disordered protein. 
Increasing the modulator size beyond the diameter of the disordered protein 
    would introduce excluded-volume and depletion effects similar to macromolecular crowding. 
For repulsive large modulators, crowding may favor compact protein assemblies 
    by increasing the free volume available to the modulator, which could lower
    the effective nucleation barrier and enhance fibrillation.
In fact, previous studies have shown that repulsive crowders larger 
    than the protein can promote fibril formation \cite{Latshaw2015}.
}

Second, our results show that when modulators 
    target disordered $\alpha$‑synuclein, they can embed within nascent oligomers 
    to result in dynamic, liquid‑like heteroassemblies (\Fig{Fig:attrS}(d)).
At intermediate modulator concentrations, 
    disordered protein assemblies persist over extended periods before converting to fibrils, consistent with a longer lag phase prior to fibril conversion.
These findings align with experiments using the 
    4554W peptide, showing delayed aggregation via selective inhibition of primary nucleation, which extends the lag phase while leaving latter growth largely unaffected \cite{Meade:jmb_432_166706_2020}.
Increased affinity towards the disordered protein, further
    favors the stabilization of disordered, dynamically exchanging clusters, which continue to sequester monomers and modulators from solution.
In this regime, one‑step nucleation becomes 
    the predominant fibrillization pathway. 
Although one‑step nucleation is faster than 
    two-step-nucleation, 
    sequestration of monomers within large, disordered assemblies reduces the available free protein, 
    limiting the monomers available for fibril growth.
Mechanistically, these pathways align with the engineered
    $\beta$-wrapin AS69 activation, which binds monomeric $\alpha$-synuclein and was shown to slow fibril elongation and block conversion of oligomeric precursors \cite{Agerschou:elife_8_2019}.
Furthermore, our results align with experiments showing 
    that the antimicrobial peptide LL-III interacts with $\alpha$-synuclein monomers and condensates to stabilize protein droplets and suppress conversion to the fibrillar state
    \cite{Oliva:chem_27_11845_2021}.

Third, $\beta$-selective modulators limit
    the growth of fibrillar structures by preventing free monomers to attach to the fibril
    tips and continue fibrillar growth.
Generally, fibrillar growth proceeds via a dock-lock-mechanism, in which free monomers first
    attach to the preformed fibril (\textit{dock}) and undergo structural rearrangements
    before they attain the fibril structure (\textit{lock}) and give rise to a new
    growth-competent surface \cite{Ilie:chemrev_119_6956_2019,Schor:jpcl_6_1076_2015,Schor:bpj_103_1296_2012,Ilie:jcp_146_115102_2017,Ilie:jctc_14_3298_2018,Jalali:jpcB_127_9759_2023,Nguyen:pnas_104_111_2007,Reddy:pnas_106_11948_2009,Straub:arpc_62_437_2011}.
At coarser resolution, this is known as stop-and-go, a mechanism in which 
    periods of active fibril elongation via monomer addition (\textit{go}) are interrupted by extended pauses in elongation (\textit{stop}) \cite{Wordehoff:jmb_427_1428_2015,Ilie:jcp_146_115102_2017,Ban:acr_39_663_2006,Ferkinghoff:pre_82_010901_2010,Hoyer:jmb_340_127_2004,Ban:jmb_344_757_2004,Pinotsi:nanolett_14_339_2014}.
Our results show that modulators targeting $\beta$-rich states 
    bind to the fibril surface, including the fibril ends, thereby sterically disfavoring
    the docking of incoming proteins.
This essentially stabilizes and extends the \textit{stop} phase of fibrillar growth.
Our results provide mechanistic insight into experimentally 
    designed peptides targeting $\beta$-rich fibrillar surfaces
    that coat the fibril surface with tunable stoichiometry \cite{Bismut:ps_34_2025}
    or stabilize the elongation-incompetent blocked state \cite{Schulz:bbaadv_4_10110_2023}.
Additionally, our results show that modulators bridge between fibrils 
    to form bundles of shorter fibrils with altered interfaces.
As such, they fundamentally change the fibril seeding properties, providing 
    a mechanistic explanation for experimental observations that bundled fibrils bridged 
    by high-affinity peptides designed from the $\alpha$-synuclein NAC core display reduced seeding ability compared to individual short fibrils \cite{Sangwan:elife_9_2020}.

\textcolor{black}{
The current study assumes purely repulsive interactions between modulators. 
Introducing attractive modulator–modulator interactions  
    would add an additional level of collective regulation. 
Even weak modulator-modulator attraction may lead to the formation of transient modulator cluster 
    and thus increase the local modulator concentration. 
Depending on the relative affinities of the modulator for the disordered and 
    $\beta$-rich protein states, this local enrichment could enhance two-step nucleation by stabilizing protein-rich intermediate assemblies, or alternatively promote sequestration into off-pathway heteroassemblies. 
Strong modulator-modulator attraction could drive modulator self-assembly or 
    phase separation into modulator-rich domains. 
Thus, by tuning modulator-modulator attraction, one could regulate the competition between 
    local concentration, phase partitioning and surface-mediated nucleation, as a natural extension of the model toward biologically heterogeneous crowded environments.
}

\textcolor{black}{
The morphing-particle protein model and the modulators represent 
    a generic framework for modulator-regulated assembly of proteins 
    that undergo a disorder-to-$\beta$-rich conformational transitions.
Here, we apply this framework to $\alpha$-synuclein as a biologically relevant example and can
    be parameterized for other fibril-forming proteins, 
    including insulin, amyloid-$\beta$, huntingtin etc. \cite{Ilie:jcp_144_1089_2016}.
Importantly, our approach is complementary to a broad family of coarse-grained amyloid-forming models, 
    ranging from patchy particles that spontaneously assemble into fibrils 
    to multibead representations of amphipathic polypeptides \cite{Ilie:chemrev_119_6956_2019,Wu:cosb_21_209_2011,Morriss-Andrews:jpcl_5_1899_2014}.
For instance, models developed by Shea and coworkers retain peptide 
    backbone and side-chain degrees of freedom, allowing sequence, peptide flexibility and 
    $\beta$-sheet propensity to influence multimerization and the resulting oligomeric and fibrillar morphologies
    \cite{Bellesia:jcp_131_111102_2009,MorrissAndrews:jcp_135_2011}.
Conformational variability is represented in the Pellarin–Caflisch model through 
    an internal degree of freedom that mediates switching between amyloid-protected and amyloid-forming states
    \cite{Pellarin:jmb_360_882_2006,Pellarin:jmb_374_917_2007}.
At a higher level of coarse graining, the model developed by Vacha and co-workers represents A$\beta$ 
    as a single patchy spherocylinder that switches between soluble and 
    $\beta$-prone binding states, thereby reproducing fibril formation \cite{Bieler:plos_8_e1002692_2012,Saric:pnas_111_17869_2014}.
Our model shares the central principle of conformation-coupled assembly yet through adaptable
    shape and interactions depending on the environment \cite{Ilie:jcp_144_1089_2016,Ilie:jcp_146_115102_2017}.
Additionally, we introduce explicit modulators with independently 
    tunable concentration, affinity and selectivity for disordered and 
    $\beta$-rich protein states, which allows us to understand how external species reshape protein phase transitions.
This minimal representation is sufficient to explain modulator-induced changes in nucleation, 
    stabilization of disordered heteroassemblies, off-pathway trapping, fibril capping and elongation inhibition. 
}

\section{Conclusions and Perspectives}
We performed coarse-grained Brownian dynamics simulations to 
    understand how modulators with distinct interaction profiles reshape $\alpha$-synuclein phase transitions. 
Our results reveal a concentration- and affinity-dependent landscape in which non-specific
    modulators stabilize dynamic, liquid-like heteroassemblies of disordered species, extending the lag phase and suppressing nucleation, while $\beta$-selective modulators accumulated on fibril surfaces, yielding shorter fibrils occasionally bundled with reduced elongation ability. 
    
These simulations go beyond experimental observations to 
    provide mechanistic insight by connecting modulator properties 
    (affinity, concentration) to outcomes (nucleation, fibril length, bundling), which may be combined with
    kinetic models \cite{Michaels:pnas_117_24251_2020,Bartolucci2026} and
    exploited in
    context of the development of novel therapeutic intervention strategies \cite{deRaffele:chemcomm_60_632_2024} and
    guide additional experimental testing.
For instance, protein sequestration within disordered heteroassemblies can suppress fibril nucleation. 
As such, the rational design of peptides, rich in hydrophobic residues
    developed to bind in the non-amyloidogenic 
    component of $\alpha$-synuclein, could stabilize intermediate species to delay or even prevent fibril formation.
Alternatively, structure-guided design of $\beta$-selective peptides with constrained backbones (via di-sulfide bonds or
    non-natural residues) can enable polymorph-specific recognition of cross-$\beta$ architectures, achieving high-affinity fibril-end blocking (via hydrogen-bonding capabilities).
Additionally, multivalent peptide conjugates linking two such $\beta$-selective constrained peptides can
    simultaneously occupy adjacent fibril surfaces
    and bridge fibrils laterally, resulting in bundled architecture, 
    whose clearance can be compared against long, individual fibrils.
Extending beyond $\alpha$-synuclein, the generic nature of the model allows the extension 
    to other polypeptides sharing the same emerging properties     
    by adjusting geometrical parameters or introducing system-specific interactions, applicable to pathological (e.g. amyloid-$\beta$, TDP-43) and non-pathological functional polypeptides (e.g. silk or oat peptides).
Such a framework can therefore inform the design of new nanomaterials with tunable and adaptable properties.
For instance, engineering stimulus-responsive materials that use modulator-controlled fibril association 
    in response to environmental
    triggers (e.g. pH, light, modulator concentration etc.) to dynamically tune mechanical properties
    or engineering nanostructures that control
    the interaction with surfaces \cite{Han:advsci_10_2023,Smith:acsnanomat_3_937_2019,Soliman:small_21_2025,Paesani_jcp_161_244905_2024,Paesani:membr,Wang:adfm_33_2023,Shao:5_46_2023}.


\section*{Supplementary Material}
The supplementary material contains additional information on the protein conformations and the number of clusters per concentration sampled during the simulations at $\constMR={}10$, $\constMS{}=5$ and $\constMR{}=5$. Additionally, it contains two movies showing the effects of non-specific and selective modulators.

\begin{acknowledgments}
 I.M.I. acknowledges support from the Sectorplan B\`{e}ta \& Techniek of the Dutch Government and the Dementia Research - Synapsis Foundation Switzerland. 
 This work has been carried out within the framework of the COST Action CA2311-SNOOPY 'Searching for Nanostructured or pOre fOrming Peptides for therapY', supported by COST (European Cooperation in Science and Technology).
\end{acknowledgments}
\newpage 
\bibliography{aipsamp.bib}

\end{document}